\documentclass[11pt,twoside]{article}
\usepackage{asp2014}

\aspSuppressVolSlug
\resetcounters

\begin{document}

\title{Improving Performance in Java: TOPCAT and STILTS}

\author{Mark~Taylor}
\affil{H.~H.~Wills Physics Laboratory, Tyndall Avenue,
       University of Bristol, UK;
       \email{m.b.taylor@bristol.ac.uk}}

\paperauthor{Mark~Taylor}{m.b.taylor@bristol.ac.uk}{0000-0002-4209-1479}{University of Bristol}{School of Physics}{Bristol}{Bristol}{BS8 1TL}{U.K.}



  
\begin{abstract}

TOPCAT and STILTS are mature Java desktop applications
for working with tabular data
that have always had a focus on efficiency for large or very
large data sets.  This paper presents some progress, experience and
lessons learned from efforts over recent years to improve performance
further by multithreading key algorithms as well as other strategies.
  
\end{abstract}

\section{Introduction}

TOPCAT, STILTS and STIL \citep{2005ASPC..347...29T,2006ASPC..351..666T}
are Java software for working with tabular
data such as source catalogues, and aim to provide astronomers with
the capability to manipulate datasets as large as possible on
modest hardware such as standard desktop or laptop machines.
How large that is depends on various factors including hardware
resources, data format and the patience of the user.
However, as a guideline to current capabilities:
the interactive GUI tool TOPCAT can comfortably be used for
exploratory analysis of tables with tens of millions of rows
and hundreds of columns,
while the command-line tool STILTS and the underlying I/O library STIL
can be used with arbitrary-sized tables; for instance STILTS
can generate a HEALPix weighted density map from a FITS file
containing the 1.8 billion row Gaia source catalogue
in $\sim$5 minutes.

Development of the software began around two decades ago,
and most of the computation was originally single-threaded.
Over that time multi-core processors have become ubiquitous,
and in the last 3--4 years effort has been expended on
trying to exploit this by multi-threading the code to
run computationally-intensive algorithms concurrently,
alongside other interventions to improve efficiency.
The changes reported here cover the period between
TOPCAT v4.6-3/STILTS 3.1-6 (May 2019) and
TOPCAT v4.8-7/STILTS v3.4-7 (October 2022).

The two main operations that are computationally intensive,
that is whose speed impacts the user experience,
are visualisation and crossmatching.

\section{Multithreading: Java Language Support}

Java provides fairly good language support for multi-threading,
though not so robust as some more modern languages like Rust;
it is still very possible to write inefficient or incorrect
concurrent Java.

Java 8 (2014) introduced {\em streams} which look like a good abstraction
for multi-threaded processing of large collections of similar items
such as table rows.
But these turned out to be hard to work with in STIL because of inflexibility
of the library interface
and opaqueness of the standard implementation; for instance
sequential evaluation is sometimes forced in circumstances
that are hard to predict, stream elements can be retained
in a way that inhibits object re-use,
and aborting stream execution is difficult.
However by following some of the patterns from Java 8 streams
and building on the lower-level classes on which they are based
(Java 7 {\tt ForkJoinPool})
a bespoke stream-like framework was implemented
to support concurrent execution of CPU-intensive algorithms.

\section{Multithreading Visualisation}

The interactive visualisation in TOPCAT (usually) operates in two stages,
which may be designated {\em caching\/} and {\em rendering}.
First the coordinate data is read from the input table,
possibly pre-processed, and {\em cached\/}
to a simplified memory-based structure.
Then, graphical frames are repeatedly {\em rendered\/}
as required from this cached data in response to user navigation actions
such as pan, zoom and rotate.
For most plot types the rendering stage is suitable for multithreading,
and since no I/O is involved this works well,
delivering near optimal acceleration at least in some regimes
(e.g.\ $\approx$8 times on 10 cores for a 100\,Mrows weighted density plot).
The caching stage can be effectively parallelised in {\em some\/} cases,
but problems arise if for instance an unpredictable selection of
the rows in the full dataset is being plotted,
so by default this is still done sequentially.

The effect of the multithreading is therefore to boost the size of dataset
that can be comfortably visualised interactively by an order of magnitude
on a typical multi-core machine.
Initiating a plot may however still take some time,
perhaps a few tens of seconds for a 100\,Mrow plot.

\section{Multithreading Cross Matching}
\label{P28:xmatchThreads}

Unlike visualisation rendering,
the cross-matching implementation works directly with
instances of the {\tt StarTable} abstraction used throughout STIL
to represent tabular data.
Multi-threading the relevant code therefore required changes to
the {\tt StarTable} interface to prepare it for concurrent access,
which was a major job (see section \ref{P28:StarTable}).
Moreover, the cross-matching algorithms involve multiple steps,
not all of which are amenable to parallelisation, so overall speedup
is harder to achieve in accordance with Amdahl's Law \citep{amdahls-law}.

Once {\tt StarTable} had been thus modified and those loops that were
suitable had been parallelised, some acceleration did result,
but overall speedup was a bit disappointing.
Profiling was undertaken to identify rate-limiting steps,
resulting in a number of other changes to improve performance.
In its current state, a parallelism of about six speeds up
end-to-end elapsed time for a typical crossmatch by around a factor
of two, though this is highly dependent on the details of the match.
Higher parallelisms yield diminishing returns in elapsed time,
so default behaviour is now to enable parallel execution with
$\leq6$ threads.

The main reason for this limited scalability is that some parts
of the algorithm cannot easily be parallelised.
Other factors resulting from fragmenting the processing also play a part:
more combination operations and more garbage collection are required,
and access to memory and disk is more scattered.

\section{StarTable Parallelisation}
\label{P28:StarTable}

STIL v4.0 (Jan 2021) introduced a number of changes required to support
parallelisation of operations based on the {\tt StarTable}
interface that serves as the basic data access abstraction in STIL.
A {\tt StarTable} may represent any source of table data, such as
a file on disk or a stream of
processing operations derived from a stored or virtual row sequence.
The main changes were introduction of the following data access methods:
\begin{itemize}
\item {\tt getRowAccess()}:
      returns a thread-safe object providing random access to table data
\item {\tt getRowSplittable()}:
      returns an object that manages splitting the row sequence into blocks
      to enable multi-threaded sequential access to table data
\end{itemize}
Since there are hundreds of subclasses of the StarTable interface
in the code, careful design and implementation of these methods
was required.
There were additionally many instances of code originally written
for a sequential context that needed to be identified and adjusted
for concurrent use.

As well as the cross-matching improvements noted in
section \ref{P28:xmatchThreads}
this enabled fairly straightforward multithreading and consequent
perfomance improvements to a few other computationally intensive
operations such as sky map generation and statistics calculations.

\section{Other Crossmatch Improvements}

Profiling the crossmatching code identified various bottlenecks
which were addressed separately to the parallelisation.
Improvements made in view of this analysis included:
\begin{itemize}
\item replacing the PixTools HEALPix library with
      the CDS HEALPix library \citep{2019ASPC..523..609P}
\item using a {\tt HashMap} instead of a {\tt TreeMap} for one data structure
\item replacing the {\tt java.lang.Math} arcsine
      with the Apache {\tt FastMath} implementation
\item improving pre-processing coverage assessment to understand
      spherical geometry
\end{itemize}
Together these changes delivered a factor of around 3 times speedup for
typical matches --- greater than the effect of the parallelisation.

\section{I/O}

Careful attention to behaviour of various I/O classes,
especially with respect to buffering, delivered significant speedups,
for instance writing to FITS and reading BINARY-encoded VOTables
were both improved by a factor of 3--4 times.
In some cases this was achieved by replacing use of the standard
library I/O classes, which are not always performant for high-volume usage,
with suitable custom alternatives.

Much of the most resource-intensive work required by these applications
is I/O-bound.
In this regime multi-threading doesn't help much and may well
degrade performance by fragmenting data access and impeding efficient
use of system caches.
This can be a major effect when reading from spinning disks for
which seek operations are expensive,
though less so when reading from solid-state disks or a hot cache.
This, and the fact that STIL abstracts data access so that application code
generally cannot distinguish memory-based from disk-based data,
make it challenging to set up reliably beneficial default
threading configurations,
for instance number of concurrent threads and
number of rows per execution block.
STILTS therefore provides options for users to experiment with concurrency.

\section{Profiling}

To identify bottlenecks a good runtime profiler is essential.
Historically most or all java CPU profiling tools have suffered from the
{\em Safepoint Bias Problem} which often gives rise to misleading
information about where CPU cycles are used.
Fortunately the
{\tt async-profiler}\footnote{\url{https://github.com/jvm-profiling-tools/async-profiler}}
application which avoids this problem
has become available in the last few years,
which was very valuable for this work.
Even so, understanding the contributions of
object allocation, garbage collection, I/O and memory access
to overall run times is still challenging.

\section{Conclusions}

Parallelising interactive visualisation has worked well;
parallelising crossmatching has taken more effort and delivered
less impressive results.
As well as the parallelisation, a number of other performance interventions
have been able to improve the user experience when working with
very large datasets in TOPCAT and STILTS, with most of the resource-intensive
operations accelerated by a factor of several over the past few years.

\bibliography{P28}


\end{document}